\documentclass[12pt]{article}
\usepackage{amsfonts,amsmath,graphicx,setspace,mathtools}
\usepackage{color,hyperref,verbatim,natbib,rotating}

\definecolor{Red}{rgb}{0.5,0,0}
\definecolor{Blue}{rgb}{0,0,0.5}
\hypersetup{%
    colorlinks = {true},
    linktocpage = {true},
    plainpages = {false},
    linkcolor = {Blue},
    citecolor = {Blue},
    urlcolor = {Red},
    pdfstartview = {XYZ null null 1.25},
    pdfpagemode = {UseOutlines},
    pdfview = {XYZ null null null}
}
\newcommand{\email}[1]{\href{mailto:#1}{\normalfont\texttt{#1}}}
\newcommand{\R}{\textsf{R}}

\newcommand{\bfalpha}{\mbox{{\boldmath $\alpha$}}}
\newcommand{\bfbeta}{\mbox{{\boldmath $\beta$}}}
\newcommand{\bfgamma}{\mbox{{\boldmath $\gamma$}}}
\newcommand{\bftheta}{\mbox{{\boldmath $\theta$}}}

\newcommand{\eps}{\varepsilon}
\newcommand{\bw}{\mbox{{\boldmath $w$}}}
\newcommand{\bv}{\mbox{{\boldmath $v$}}}
\newcommand{\by}{\mbox{{\boldmath $y$}}}
\newcommand{\bx}{\mbox{{\boldmath $x$}}}

\newcommand{\bbI}{\mbox{\textbf I}}
\newcommand{\bz}{\mbox{{\boldmath $z$}}}

\newcommand{\bK}{\mbox{{\boldmath $K$}}}
\newcommand{\bb}{\mbox{{\boldmath $b$}}}

\newcommand{\bB}{\mbox{{\boldmath $B$}}}

\newcommand{\bD}{\mbox{{\boldmath $D$}}}
\newcommand{\dcl}{\prescript{}{\mbox{cv}}{\mbox{DCL}}}
\newcommand{\dclHat}{\prescript{}{\mbox{cv}}{\mbox{D}\widehat{\mbox{C}}\mbox{L}}}
\def\argmax{\mathop{\rm arg\,max}}

\bibpunct{(}{)}{;}{a}{}{,}

\begin{document}

{\vspace*{1cm}}

\begin{center}
\Large \bf Personalized Screening Intervals for Biomarkers using Joint Models for Longitudinal and Survival Data
\end{center}

\vspace{0.5cm}
\begin{center}
{\large Dimitris Rizopoulos$^{1,*}$, Jeremy M.G. Taylor$^2$, Joost van Rosmalen$^{1}$, Ewout W. Steyerberg$^3$, Johanna J.M. Takkenberg$^4$}\footnote{$^*$Correspondance at: Department of Biostatistics, Erasmus University Medical Center, PO Box 2040, 3000 CA Rotterdam, the Netherlands. E-mail address: \email{d.rizopoulos@erasmusmc.nl}.}\\
$^{1}$Department of Biostatistics, Erasmus University Medical Center\\
$^{2}$Department of Biostatistics, University of Michigan - Ann Arbor, USA\\
$^{3}$Department of Public Health, Erasmus University Medical Center, the Netherlands\\
$^{4}$Department of Cardiothoracic Surgery, Erasmus University Medical Center, the Netherlands
\end{center}
\vspace{0.6cm}


\begin{spacing}{1}
\noindent {\bf Abstract}\\
Screening and surveillance are routinely used in medicine for early detection
of disease and close monitoring of progression. Biomarkers are one of the primarily tools used for these tasks, but their successful translation to clinical practice is closely linked to their ability to accurately predict clinical endpoints during follow-up. Motivated by a study of patients who received a human tissue valve in the aortic position, in this work we are interested in optimizing and personalizing screening intervals for longitudinal biomarker measurements. Our aim in this paper is twofold: First, to appropriately select the model to use at time $t$, the time point the patient was still event-free, and second, based on this model to select the optimal time point $u > t$ to plan the next measurement. To achieve these two goals we develop measures based on information theory quantities that assess the information we gain for the conditional survival process given the history of the subject that includes both baseline information and his/her accumulated longitudinal measurements.\\\\
\noindent {\it Keywords:} Biomarkers, Decision making, Personalized medicine, Random effects.
\end{spacing}


\section{Introduction} \label{Sec:Intro}
\begin{figure}[!h]
\includegraphics[width = \textwidth]{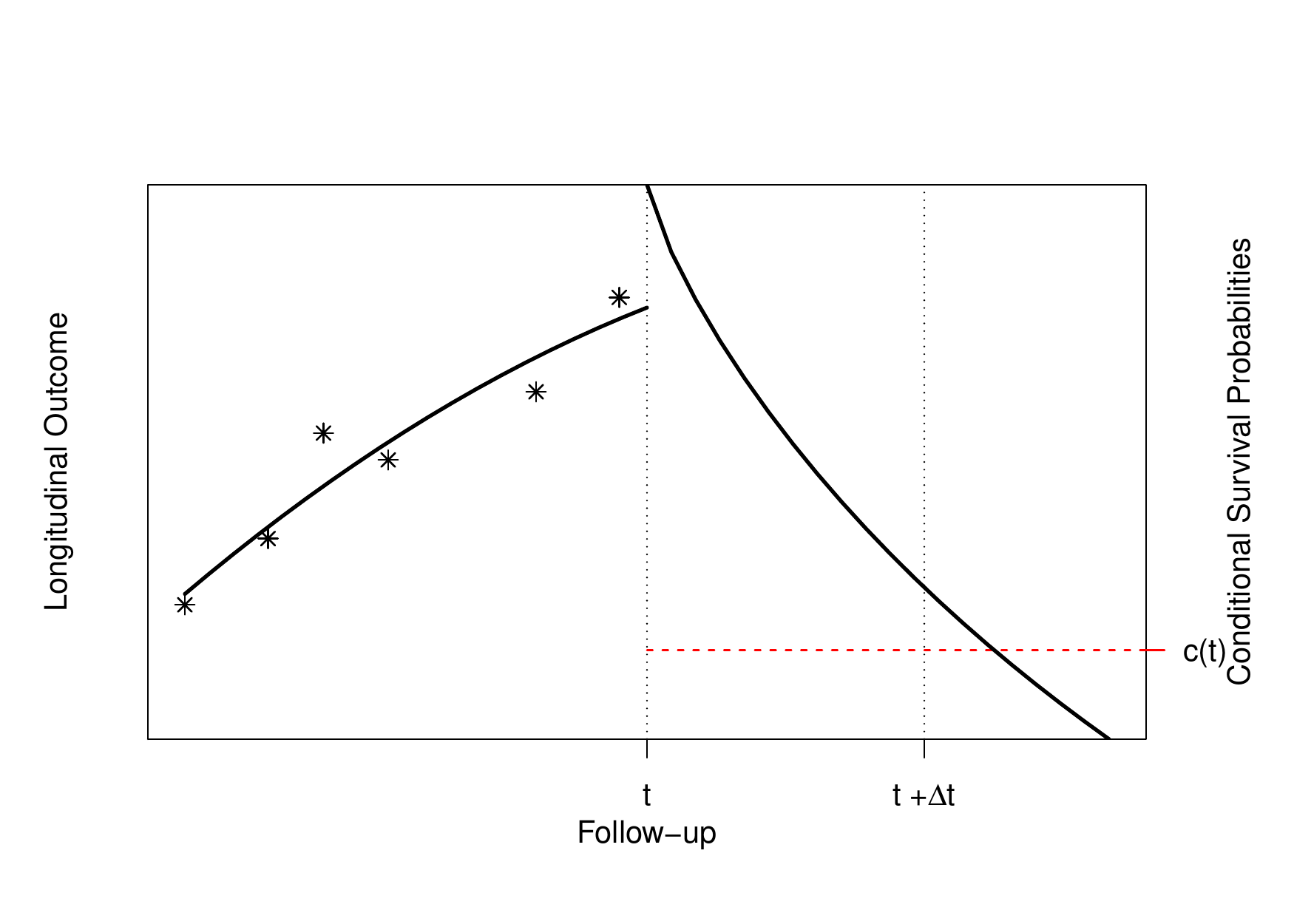}
\caption{Illustration of a patient's longitudinal profile up to time $t$, and conditional survival function for whom we wish to plan an extra measurement. Physicians are interested in events occurring within the time interval $(t, t + \Delta t]$. Constant $c(t)$ defines the threshold above which the physician requires extra information before making a decision.} \label{Fig:NextVisit}
\end{figure}

Decision making in medicine has become increasingly complex for patients and practitioners. This has resulted from factors such as the shift away from physician authority toward shared decision making, unfiltered information on the Internet, new technology providing additional data, numerous treatment options with associated risks and benefits, and results from new clinical studies. Within this context medical screening procedures are routinely performed for several diseases. Most prominent examples can be found in cancer research, where, for example, mammographic screening is performed for the detection of breast cancer in asymptomatic patients, and prostate-specific antigen (PSA) levels are used to monitor the progression of prostate cancer in men who had already been diagnosed with the disease. In general, the aim of screening procedures is to optimize the benefits, i.e., early detection of disease or deterioration of the condition of a patient, while also balancing the respective costs. In this paper we are interested in optimizing screening intervals for asymptomatic or symptomatic patients that are followed-up prospectively (the latter possibly after a medical procedure). The setting we focus on is better explained via Figure~\ref{Fig:NextVisit}.
More specifically, this figure depicts a hypothetical patient who has been followed-up up to time point $t$, and has provided a set of longitudinal biomarker measurements. Using this information, physicians are interested in events occurring within a medically relevant time interval $(t, t + \Delta t]$. If the probability of surviving beyond $t + \Delta t$ is lower than the pre-specified constant $c(t)$, then the physician is supposed to immediately take action to improve the prospects for this patient. However, when the chance of surviving until $t + \Delta t$ is higher than $c(t)$, then the physician may not initiate treatment or require an invasive test, and instead wishes to gain a better understanding of the disease progression using additional biomarker measurements. The decision to request an additional measurement requires careful weighing of the potential benefits and costs to the patient and health care system. Hence, we are interested in carefully planning the timing of the next measurement in order to maximize the information we can gain on disease progression, and at the same time minimize costs and patient burden.

Statistical methods for optimizing medical screening strategies have flourished in the past years. These methods have been primarily based on Markov and multistate models (\citealp{sonnenberg.beck:93, parmigiani:93, lee.zelen:98, parmigiani:98, parmigiani:99, parmigiani:02, schaefer.et.al:04, alagoz.et.al:10}). The appealing feature of these models is that they can simultaneously handle different medical options and select the one that provides the most gains at an acceptable cost (cost-effectiveness). Here we will instead work under the framework of joint models for longitudinal and time-to-event data (\citealp{henderson.et.al:00, tsiatis.davidian:04, rizopoulos:12}). An advantageous feature of joint models is that they utilize random effects and therefore have an inherent subject-specific nature. This allows one to better tailor decisions to individual patients, i.e., personalize screening, rather than using the same screening rule for all. In addition, contrary to the majority of screening approaches that are based on a dichotomized version of the last available biomarker measurement to indicate the need for a further medical procedure (e.g., PSA greater than 4.0 ng/mL), joint models utilize the whole longitudinal history to reach a decision without requiring dichotomization of the longitudinal outcome. Our methodological developments couple joint models with information theory quantities, following similar arguments as in \citet{commenges.et.al:12}. In particular, our aim in this paper is twofold: First, to appropriately select the joint model to use at time $t$, i.e. the time point the subject of interest is still event-free, and second, based on this model to select the optimal time point $u > t$ to plan the next measurement. We measure optimality by the amount of information gained for the survival process given the history of the subject of interest that includes both baseline information and his/her accumulated longitudinal measurements. Depending on the characteristics of the disease under study, the physician may request that $u$ lies within the interval $(t, t + \Delta t]$ or set another upper time limit after which he/she is not willing to wait before obtaining the next measurement.

The motivation for this research comes from an aortic valve study conducted by the Department of Cardio-Thoracic Surgery of the Erasmus Medical Center in the Netherlands. This study includes 285 patients who received a human tissue valve in the aortic position in the hospital from 1987 until 2008 (\citealp{bekkers.et.al:11}). Aortic allograft implantation has been widely used for a variety of aortic valve or aortic root diseases. Major advantages ascribed to allografts are the excellent hemodynamic characteristics as a valve substitute; the low rate of thrombo-embolic complications, and, therefore, absence of the need for anticoagulant treatment and the resistance to endocarditis. A major disadvantage of using human tissue valves, however is the susceptibility to degeneration and the concomitant need for re-interventions. The durability of a cryopreserved aortic allograft is age-dependent, leading to a high lifetime risk of re-operation, especially for young patients. Re-operations on the aortic root are complex, with substantial re-operative risks, and mortality rates in the range of 4--12\%. A timely well-planned re-operation has lower mortality risks compared to urgent and re-operative surgery. It is therefore of great interest for cardiologists and cardio-thoracic surgeons to plan in the best possible manner echocardiographic assessments that will inform them about the future prospect of a patient with a human tissue valve in order to optimize medical care, carefully plan re-operation and minimize valve-related morbidity and mortality.

The rest of the paper is organized as follows. Section~\ref{Sec:JM} briefly presents the joint modeling framework that sets the basis for our later methodological developments. Section~\ref{Sec:Model-Choice} presents the information theory quantities that can be used to select the joint model at a particular follow-up time, and Section~\ref{Sec:Info} the corresponding quantities for planning the next longitudinal measurement. Section~\ref{Sec:AoValv} illustrates the use of these information measures in the Aortic Valve study, and Section~\ref{Sec:Disc} concludes the paper.


\section{Joint Model Specification} \label{Sec:JM}
In this section we present a general definition of the framework of joint models for longitudinal and survival data that will be used later on for planning the optimal visit schedule. Let $\mathcal D_n = \{T_i, \delta_i, \by_i; i = 1, \ldots, n\}$ denote a sample from the target population, where $T_i^*$ denotes the true event time for the $i$-th subject, $C_i$ the censoring time, $T_i = \min(T_i^*, C_i)$ the corresponding observed event time, and $\delta_i = I(T_i^* \leq C_i)$ the event indicator, with $I(\cdot)$ being the indicator function that takes the value 1 when $T_i^* \leq C_i$, and 0 otherwise. In addition, we let $\by_i$ denote the $n_i \times 1$ longitudinal response vector for the $i$-th subject, with element $y_{il}$ denoting the value of the longitudinal outcome taken at time point $t_{il}$, $l = 1, \ldots, n_i$.

To accommodate different types of longitudinal responses in a unified framework, we postulate a generalized linear mixed effects model. In particular, the conditional distribution of $\by_i$ given a vector of random effects $\bb_i$ is assumed to be a member of the exponential family, with linear predictor given by
\begin{equation}
k \bigl [ E \{ y_i(t) \mid \bb_i \} \bigr ] = \eta_i(t) =
\bx_i^\top(t) \bfbeta + \bz_i^\top(t) \bb_i, \label{Eq:MixedModel}
\end{equation}
where $k(\cdot)$ denotes a known one-to-one monotonic link function, and $y_i(t)$ denotes the value of the longitudinal outcome for the $i$-th subject at time point $t$, $\bx_i(t)$ and $\bz_i(t)$ denote the time-dependent design vectors for the fixed-effects $\bfbeta$ and for the random effects $\bb_i$, respectively. The random effects are assumed to follow a multivariate normal distribution with mean zero and variance-covariance matrix $\bD$. For the survival process, we assume that the risk for an event depends on a function of the subject-specific linear predictor $\eta_i(t)$ and/or the random effects. More specifically, we have
\begin{eqnarray}
\nonumber h_i (t \mid \mathcal H_i(t), \bw_i) & = &
\lim_{\Delta t \rightarrow 0} \Pr \{ t \leq T_i^* < t + \Delta t \mid T_i^* \geq t, \mathcal H_i(t),
\bw_i \} \big / \Delta t\\
& = & h_0(t) \exp \bigl  [\bfgamma^\top
\bw_i + f \{\eta_i(t), \bb_i, \bfalpha \} \bigr] , \quad t > 0, \label{Eq:Surv-RR}
\end{eqnarray}
where $\mathcal H_i(t) = \{ \eta_i(s), 0 \leq s < t \}$ denotes the history of the underlying longitudinal process up to $t$, $h_0(\cdot)$ denotes the baseline hazard function, $\bw_i$ is a vector of baseline covariates with corresponding regression coefficients $\bfgamma$. Function $f(\cdot)$, parameterized by vector $\bfalpha$, specifies which components/features of the longitudinal outcome process are included in the linear predictor of the relative risk model. Some examples, motivated by the literature \citep{brown:09, rizopoulos.ghosh:11, rizopoulos:12, taylor.et.al:13, rizopoulos.et.al:14}, are:
\begin{eqnarray*}
f \{\mathcal H_i(t), \bb_i, \bfalpha \} & = & \alpha \eta_i(t),\\
f \{\mathcal H_i(t), \bb_i, \bfalpha \} & = & \alpha_1 \eta_i(t) + \alpha_2 \eta_i'(t),
\mbox{ with } \eta_i'(t) = \frac{d\eta_i(t)}{dt},\\
f \{\mathcal H_i(t), \bb_i, \bfalpha \} & = & \alpha \int_0^t \eta_i(s) \, ds,\\
f \{\mathcal H_i(t), \bb_i, \bfalpha \} & = & \bfalpha^\top \bb_i.
\end{eqnarray*}
These formulations of $f(\cdot)$ postulate that the hazard of an event at time $t$ may be associated with the underlying level of the biomarker at the same time point, the slope of the longitudinal profile at $t$, the accumulated longitudinal process up to $t$, or the random effects alone. Finally,  the baseline hazard function $h_0(\cdot)$ is modeled flexibly using a B-splines approach, i.e.,
\begin{equation}
\log h_0(t) = \gamma_{h_0,0} +
\sum \limits_{q = 1}^Q \gamma_{h_0,q} B_q(t, \bv),
\label{Eq:BaseHaz}
\end{equation}
where $B_q(t, \bv)$ denotes the $q$-th basis function of a B-spline with knots $v_1, \ldots, v_Q$ and $\bfgamma_{h_0}$ the vector of spline coefficients. To avoid the task of choosing the appropriate number and position of the knots, we include a relatively high number of knots (e.g., 15 to 20) and appropriately penalize the B-spline regression coefficients $\bfgamma_{h_0}$ for smoothness using the differences penalty \citep{eilers.marx:96}.

For the estimation of joint model's parameters we use a Bayesian approach based on Markov chain Monte Carlo (MCMC) algorithms. The expression for the posterior distribution of the model parameters given the observed data is derived under the assumptions that given the random effects, both the longitudinal and event time process are assumed independent, and the longitudinal responses of each subject are assumed independent. Formally we have,
\begin{eqnarray}
p(\by_i, T_i, \delta_i \mid \bb_i, \bftheta) & = & p(\by_i \mid \bb_i, \bftheta) \;
p(T_i, \delta_i \mid \bb_i, \bftheta), \label{Eq:CondInd-I}\\
p(\by_i \mid \bb_i, \bftheta) & = & \prod_l p ( y_{il} \mid \bb_i, \bftheta ),
\label{Eq:CondInd-II}
\end{eqnarray}
where $\bftheta$ denotes the full parameter vector, and $p(\cdot)$ denotes an appropriate probability density function. Under these assumptions the posterior distribution is analogous to:
\begin{eqnarray}
p(\bftheta, \bb) \propto \prod \limits_{i = 1}^n \prod \limits_{l = 1}^{n_i}
p (y_{il} \mid \bb_i, \bftheta) \; p(T_i, \delta_i \mid \bb_i, \bftheta) \;
p(\bb_i \mid \bftheta) \; p(\bftheta), \label{Eq:FullPost}
\end{eqnarray}
where
\[
p (y_{il} \mid \bb_i, \bftheta) = \exp \biggl \{ \Bigl [y_{il} \psi_{il}(\bb_i) -
c\{\psi_{il}(\bb_i)\} \Bigr ] \Big / a(\varphi) - d(y_{il}, \varphi) \biggr \},
\]
with $\psi_{il}(\bb_i)$ and $\varphi$ denoting the natural and dispersion parameters in the exponential family, respectively, $c(\cdot)$, $a(\cdot)$, and $d(\cdot)$ are known functions specifying the member of the exponential family, and for the survival part
\[
p(T_i, \delta_i \mid \bb_i, \bftheta) = h_i(T_i \mid \mathcal H_i(T_i, \bb_i))^{\delta_i}
\exp \Bigl \{- \int_0^{T_i} h_i(s \mid \mathcal H_i(s, \bb_i) ) \; ds \Bigl\},
\]
with $h_i(\cdot)$ given by \eqref{Eq:Surv-RR}. The integral in the definition of the survival function
\begin{equation}
S_i(t \mid \mathcal H_i(t), \bb_i, \bw_i) = \exp \Bigl \{- \int_0^t h_0(s)
\exp \bigl  [\bfgamma^\top \bw_i + f \{\eta_i(s), \bb_i, \bfalpha \} \bigr] ds \Bigr \},
\label{Eq:SurvivalFun}
\end{equation}
does not have a closed-form solution, and thus a numerical method must be employed for its evaluation. Standard options are the Gauss-Kronrod and Gauss-Legendre quadrature rule.

The penalized version of the B-spline approximation to the baseline hazard can be fitted by specifying for $\bfgamma_{h_0}$ the improper prior \citep{lang.brezger:04}:
\[
p(\bfgamma_{h_0} \mid \tau_h) \propto \tau_h^{\rho(K)/2}\exp \Bigl (-\frac{\tau_{h}}{2}
\bfgamma_{h_0}^\top \bK \bfgamma_{h_0} \Bigr ),
\]
where $\tau_h$ is the smoothing parameter that takes a $\mbox{Gamma}(1, \tau_{h\delta} )$ prior distribution, with a hyper-prior $\tau_{h\delta} \sim \mbox{Gamma}(10^{-3}, 10^{-3})$, which ensures a proper posterior distribution for $\bfgamma_{h_0}$ \citep{jullion.lambert:07}, $\bK = \Delta_r^\top \Delta_r + 10^{-6}\bbI$, with $\Delta_r$ denoting the $r$-th difference penalty matrix, and $\rho(\bK)$ denotes the rank of $\bK$.


\section{Choosing the Model to Use at Time $t$} \label{Sec:Model-Choice}
As motivated in Section~\ref{Sec:Intro}, our aim is to decide the appropriate time point to plan the next measurement for a new subject $j$ from the same population (i.e., with $j \not \in \{i = 1, \dots, n\}$), whose survival chance at a particular time point $t$ lies within the interval $[c(t), 1]$ (see Figure~\ref{Fig:NextVisit}). Recent research in the field of joint models \citep{yu.et.al:08, proust-lima.taylor:09, rizopoulos:11, rizopoulos:12, taylor.et.al:13} has shown how such conditional survival probabilities can be computed in order to identify for which subjects we need to proceed in planning the next measurement. In particular, the fact that biomarker measurements have been recorded up to $t$, implies that subject $j$ was event-free up to this time point, and therefore we are interested in:
\begin{equation}
\pi_j(u \mid t) = \Pr \bigl \{ T_j^* > u \mid T_j^* > t, \mathcal Y_j(t), \mathcal D_n \bigr \},
\label{Eq:conSurv}
\end{equation}
where $\mathcal Y_j(t) = \{ y_j(t_{jl}); 0 \leq t_{jl} \leq t, l = 1, \ldots, n_j \}$ denotes the longitudinal measurements of subject $j$ taken before $t$, and $\mathcal D_n = \{T_i, \delta_i, \by_i; i = 1, \dots, n\}$ denotes the original dataset.

As we have seen in Section~\ref{Sec:JM}, in general, defining a joint model entails choosing an appropriate model for the longitudinal outcome (i.e., baseline covariates and functional form of the time effect), an appropriate model for the time-to-event (i.e., baseline covariates), and how to link the two outcomes using function $f(\cdot)$. Hence, a relevant question is which model to use for deciding when to plan the next measurement of subject $j$. Standard approaches for model selection within the Bayesian framework include the deviance information criterion (DIC) and (pseudo-) Bayes factors. However, a potential problem with these methods is that they provide an overall assessment of a model's predictive ability, whereas we are interested in the model that best predicts future events given the fact that subject $j$ was event-free up to time point $t$. To identify the model to use at $t$ we follow similar arguments as in \citet{commenges.et.al:12}, and focus on the conditional density function of the survival outcome given the longitudinal responses and survival up to $t$. More specifically, we let $\mathcal M = \{ M_1, \ldots, M_K\}$ denote a set of $K$ joint models fitted to the original dataset $\mathcal D_n$. Due to the fact that the choice of the optimal model in $\mathcal M$ is based on a finite sample, an important issue we need to address is overfitting. Hence, to obtain an objective estimate of the predictive ability of each model we will use the cross-validatory posterior predictive conditional density of the survival outcome, which for model $M_k$ is defined as $p \bigl \{ T_i^* \mid T_i^* > t, \mathcal Y_i(t), \mathcal D_{n \setminus i}, M_k \bigr \}$, where $\mathcal D_{n \setminus i} = \{T_{i'}, \delta_{i'}, \by_{i'}; i' = 1, \dots, i - 1, i + 1, \ldots, n\}$ denotes the version of the dataset that excludes the responses of the $i$-th subject. Similarly we can define the same distribution under model $M^*$, which denotes the model under which the data have been generated ($M^*$ is not required to be in $\mathcal M$). Using conventional quantities of information theory \citep{cover.thomas:91}, we select the model $M_k$ in the set $\mathcal M$ that minimizes the cross-entropy:
\[
\mbox{CE}_k(t) = E \biggl \{ - \log \Bigl [ p \bigl \{ T_i^* \mid T_i^* > t, \mathcal Y_i(t), \mathcal D_{n \setminus i}, M_k \bigr \} \Bigr ] \biggr \}, \label{Eq:ECE}
\]
where the expectation is taken with respect to $p(T_i^* \mid T_i^* > t, \mathcal Y_i(t), \mathcal D_{n \setminus i}, M^*)$. An estimate of $\mbox{CE}_k(t)$ that accounts for censoring can be obtained using the available information in the sample at hand. We term this estimate the \emph{cross-validated Dynamic Conditional Likelihood}:
\begin{equation}
\dcl_k(t) = \frac{1}{n_t} \sum \limits_{i = 1}^n - I(T_i > t) \log p \bigl \{ T_i, \delta_i \mid T_i > t, \mathcal Y_i(t), \mathcal D_{n \setminus i}, M_k \bigr \}, \label{Eq:cvDCL}
\end{equation}
where $n_t = \sum_i I(T_i > t)$. In turn, an estimate of $p \bigl \{ T_i, \delta_i \mid T_i^* > t, \mathcal Y_i(t), \mathcal D_{n \setminus i}, M_k \bigr \}$ can be obtained from the MCMC sample of model $M_k$ by utilizing a similar computation as for the conditional predictive ordinate statistic \citep[condition on $M_k$ is dropped in the following expressions but is assumed]{chen.et.al:08}:
\begin{eqnarray}
\nonumber \lefteqn{\Bigl [ p \bigl \{ T_i, \delta_i \mid T_i > t, \mathcal Y_i(t), \mathcal D_{n \setminus i} \bigr \} \Bigr ]^{-1} = \frac{p \{ \mathcal D_{n \setminus i}, T_i > t, \mathcal Y_i(t)\}}{p \{T_i, \delta_i, T_i > t, \mathcal Y_i(t), \mathcal D_{n \setminus i} \} }}\\ \nonumber &&\\
& = & \nonumber \int \frac{p(\mathcal D_{n \setminus i} \mid \bftheta) \, p \{T_i > t, \mathcal Y_i(t) \mid \bftheta \} \, p(\bftheta)}{p(\mathcal D_n)} \, d\bftheta = \int \frac{p\{T_i > t, \mathcal Y_i(t) \mid \bftheta\}}{p\{T_i, \delta_i, \mathcal Y_i(t) \mid \bftheta\}} \frac{p(\mathcal D_n \mid \bftheta) \, p(\bftheta)}{p(\mathcal D_n)} \, d\bftheta\\ \nonumber &&\\
& = & \int \frac{1}{p\{T_i, \delta_i \mid T_i > t, \mathcal Y_i(t), \bftheta \}} \, p(\bftheta \mid \mathcal D_n) \, d\bftheta, \label{Eq:cvDCL-est}
\end{eqnarray}
where $\bftheta$ includes here both the parameters of the joint model $\bftheta$, as defined in Section~\ref{Sec:JM}, and the random effects. Hence, based on the MCMC sample from $[\bftheta \mid \mathcal D_n]$ and identity \eqref{Eq:cvDCL-est} we can obtain a Monte Carlo estimate of \eqref{Eq:cvDCL} as
\begin{equation}
\dclHat(t) = \frac{1}{n_t} \sum \limits_{i = 1}^n I(T_i > t) \log \biggl ( \frac{1}{G} \sum \limits_{g = 1}^G \frac{1}{p \bigl \{ T_i, \delta_i \mid T_i > t, \mathcal Y_i(t), \bftheta^{(g)} \bigr \} } \biggr ), \label{Eq:cvDCL-MCMC}
\end{equation}
where $\bftheta^{(g)}$ denotes here the $g$-th realization from the MCMC sample $\{\bftheta^{(g)}; g = 1, \ldots, G\}$. Calculation of $p \bigl \{ T_i, \delta_i \mid T_i > t, \mathcal Y_i(t), \bftheta^{(g)} \bigr \}$ is based on the conditional independence assumptions \eqref{Eq:CondInd-I} and \eqref{Eq:CondInd-II}. Due to space limitations and the fact that similar calculations will be required in the following section we do not show them in detail here.


\section{Scheduling Patterns based on Information Theory} \label{Sec:Info}
\subsection{Planning of the Next Measurement} \label{Sec:Info-NextVisit}
Having chosen the appropriate model to use at time $t$, our aim next is to decide when to plan the following measurement of subject $j$. Similarly to the previous section, the quantity that we will use for making this decision is the posterior predictive conditional distribution $p\{T_j^* \mid T_j^* > t, \mathcal Y_j(t), \mathcal D_n\}$. This distribution again has a cross-validatory flavor because $j \not\in \{i = 1, \ldots, n\}$. We let $y_j(u)$ denote the measurement of the longitudinal outcome at the future time point $u > t$. For the selection of the optimal $u$ we would like to maximize the information we gain by measuring $y_j(u)$ at this time point, provided that the patient was still event-free up to $u$. If the event occurs before time $u$ for subject $j$, then there is no gain in information. Using concepts from Bayesian optimal designs \citep{verdinelli.kadane:92, clyde.chaloner:96}, this can be formally expressed using the utility function (conditioning on baseline covariates $\bw_j$ is assumed in the following expressions but is dropped for notational simplicity):
\begin{equation}
\mbox{U}(u \mid t) = E \biggl \{ \lambda_1 \log \frac{p \bigl (T_j^* \mid T_j^* > u, \bigl \{ \mathcal Y_j(t), y_j(u) \bigr \}, \mathcal D_n \bigr )}{p\{T_j^* \mid T_j^* > u, \mathcal Y_j(t), \mathcal D_n\}} + \lambda_2 I(T_j^* > u) \biggr \}, \label{Eq:Util}
\end{equation}
where the expectation is taken with respect to the joint predictive distribution $[T_j^*, y_j(u) \mid T_j^* > t, \mathcal Y_j(t), \mathcal D_n]$. Making the connection to the information theory concepts we used in Section~\ref{Sec:Model-Choice}, the first term in \eqref{Eq:Util} is the expected Kullback-Leibler divergence between the posterior predictive conditional distributions with and without this extra measurement, namely
\begin{eqnarray}
\lefteqn{\mbox{EKL}(u \mid t) = E_{Y} \Biggl [ E_{T^* \mid Y} \biggl \{ \log \frac{p \bigl (T_j^* \mid T_j^* > u, \bigl \{ \mathcal Y_j(t), y_j(u) \bigr \}, \mathcal D_n \bigr )}{p\{T_j^* \mid T_j^* > u, \mathcal Y_j(t), \mathcal D_n\}} \biggr \} \Biggr ] } \label{Eq:KL} \\
& = & \nonumber \int \biggl \{ \int \log \frac{p \bigl (T_j^* \mid T_j^* > u, \bigl \{ \mathcal Y_j(t), y_j(u) \bigr \}, \mathcal D_n \bigr )}{p\{T_j^* \mid T_j^* > u, \mathcal Y_j(t), \mathcal D_n\}} \, p \bigl ( T_j^* \mid T_j^* > t, \bigl \{ \mathcal Y_j(t), y_j(u) \bigr \}, \mathcal D_n \bigr ) \, dT_j^*
\biggr \}\\
&& \nonumber \hspace*{7cm} \times \, p\{y_j(u) \mid T_j^* > t, \mathcal Y_j(t), \mathcal D_n\} \, dy_j(u).
\end{eqnarray}
Heuristically, for $u$ close to $t$, in general, we would expect minimal information gain. Analogously, as $u$ moves further away from $t$ we would expect to gain information, but the value of $u$ that maximizes this information gain will depend on the characteristics of the subject's projected longitudinal profile. However, the `cost' of waiting up to time $u$, is the fact that we also inevitably increase the risk that the patient will experience the event. Because we would not like to wait up to a point that it would be late for the physician to intervene (e.g., the event already happened), we need to counterbalance these two aspects, which motivates including the second term in \eqref{Eq:Util}. The expectation with respect to this second term is the conditional survival probability $\pi_j(u \mid t)$ introduced in \eqref{Eq:conSurv}. Hence, \eqref{Eq:Util} becomes
\[
\mbox{U}(u \mid t) = \lambda_1 \mbox{EKL}(u \mid t) + \lambda_2 \pi_j(u \mid t).
\]
The purpose of the nonnegative constants $\lambda_1$ and $\lambda_2$ is to weigh the contribution of the cumulative risk as opposed to the information gain, and to take into account that these two terms have different units. In practical applications, elicitation of these constants for trading information units with probabilities can be difficult. However, using the equivalence between compound and constrained optimal designs \citep{cook.wong:94, clyde.chaloner:96}, it can be shown that for any $\lambda_1$ and $\lambda_2$, there exists a constant $\kappa \in [0, 1]$ for which maximization of \eqref{Eq:Util} is equivalent to maximization of $\mbox{EKL}(u \mid t)$ subject to the constraint that $\pi_j(u \mid t) \geq \kappa$. Under this equivalent formulation elicitation of $\kappa$ is relatively easier; for example, a logical choice is to set $\kappa$ equal to the constant $c(t)$, introduced in Figure~\ref{Fig:NextVisit}, for deciding whether the physician is supposed to intervene or not. In turn, a choice of $c(t)$ can be based on medical grounds or on time-dependent receiver operating characteristic (ROC) analysis. Then, the optimal value of $u$ can be found by maximizing $\mbox{EKL}(u \mid t)$ in the interval $(t, t^{up}]$, where $t^{up} = \{u: \pi_j(u \mid t) = \kappa\}$. In practice, it also may be reasonable to assume that there is an upper limit $t^{max}$ of the time interval the physician is willing to wait to obtain the next measurement, in which case $t^{up} = \min\{u: \pi_j(u \mid t) = \kappa, \, t^{max}\}$. The advantageous feature of this approach is that the optimal $u$ is found by conditioning on both the baseline characteristics $\bw_j$ and accumulated longitudinal information $\mathcal Y_j(t)$ of subject $j$. This allows one to better tailor screening to the subject's individual characteristics compared to approaches that only utilize the baseline covariates and the last available longitudinal measurement (Markov assumption).


\subsection{Estimation} \label{Sec:Info-Est}
Estimation of \eqref{Eq:Util} proceeds by suitably utilizing the conditional independence assumptions \eqref{Eq:CondInd-I} and \eqref{Eq:CondInd-II}. More specifically, the density of the joint distribution $[T_j^*, y_j(u) \mid T_j^* > t, \mathcal Y_j(t), \mathcal D_n]$ can be written as the product of the predictive distributions,
\begin{equation*}
p \bigl (T_j^* \mid  T_j^* > t, \bigl \{ \mathcal Y_j(t), y_j(u) \bigr \}, \mathcal D_n \bigr ) = \int p \bigl (T_j^* \mid T_j^* > t, \bigl \{ \mathcal Y_j(t), y_j(u) \bigr \}, \bftheta \bigr ) \, p(\bftheta \mid \mathcal D_n) \, d\bftheta, \label{Eq:TgvYtYu}
\end{equation*}
and,
\begin{equation*}
p\{y_j(u) \mid T_j^* > t, \mathcal Y_j(t), \mathcal D_n\} = \int p \{ y_j(u) \mid T_j^* > t, \mathcal Y_j(t), \bftheta\} \, p(\bftheta \mid \mathcal D_n) \, d\bftheta. \label{Eq:postPred-Long}
\end{equation*}
Analogously, the first term in the right-hand side of each \eqref{Eq:TgvYtYu} and \eqref{Eq:postPred-Long} can be written as expectations over the corresponding posterior distributions of the random effects of subject $j$, i.e.,
\begin{equation*}
p\{y_j(u) \mid T_j^* > t, \mathcal Y_j(t), \bftheta\} = \int p\{y_j(u) \mid \bb_j, \bftheta\} \, p\{ \bb_j \mid T_j^* > t, \mathcal Y_j(t), \bftheta\} \, d\bb_j, \label{Eq:YugvYttheta}
\end{equation*}
and,
\begin{eqnarray*}
\nonumber \lefteqn{p \bigl (T_j^* \mid T_j^* > t, \bigl \{ \mathcal Y_j(t), y_j(u) \bigr \}, \bftheta \bigr )}\\
& = & \int p (T_j^* \mid T_j^* > t, \bb_j, \bftheta ) \, p\{\bb_j \mid T_j^* > t, \bigl \{ \mathcal Y_j(t), y_j(u) \bigr \}, \bftheta\} \, d\bb_j, \label{Eq:TgvYtYutheta}
\end{eqnarray*}
respectively. In a similar manner, the density of the predictive distribution $[T_j^* \mid T_j^* > u, \bigl \{ \mathcal Y_j(t), y_j(u) \bigr \}, \mathcal D_n]$ is written as:
\begin{eqnarray*}
\lefteqn{p(T_j^* \mid T_j^* > u, \bigl \{ \mathcal Y_j(t), y_j(u) \bigr \}, \mathcal D_n)}\\
& = & \int \int p(T_j^* \mid T_j^* > u, \bb_j, \bftheta) \, p(\bb_j \mid T_j^* > u, \bigl \{ \mathcal Y_j(t), y_j(u) \bigr \}, \bftheta) \, p(\bftheta \mid \mathcal D_n) \, d\bb_j d\bftheta,
\end{eqnarray*}
with
\begin{eqnarray*}
p(T_j^* \mid T_j^* > u, \bb_j, \bftheta) = \frac{h_j \bigl (T_j^* \mid \mathcal H_j \bigl (t, \bb_j, \bftheta \bigr)\bigr) \exp \biggl \{- \displaystyle \int_0^{T_j^*} h_j \bigl (s \mid \mathcal H_j \bigl (t, \bb_j, \bftheta \bigr)\bigr) \, ds\biggr \}}{\exp \biggl \{- \displaystyle \int_0^u h_j \bigl (s \mid \mathcal H_j \bigl (t, \bb_j, \bftheta \bigr)\bigr) \, ds\biggr \}},
\end{eqnarray*}
and the hazard function is given by \eqref{Eq:Surv-RR}, in which we have explicitly noted that the history of the longitudinal process $\mathcal H_j \bigl (t, \bb_j, \bftheta \bigr)$ is a function of both the random effects and the parameters, because it is estimated from the mixed model \eqref{Eq:MixedModel}. By combining the above equations we can construct a Monte Carlo simulation scheme to obtain an estimate of \eqref{Eq:Util} for a specific value of $u$. This scheme contains the following steps that should be repeated $q = 1, \ldots, Q$ times:
\begin{itemize}
\item[Step 1:] Simulate $\widetilde \bftheta$, $\ddot \bftheta$, and $\{\breve \bftheta^{(\ell)}, \ell = 1, \ldots, L\}$ independently from the posterior distribution of the parameters $[\bftheta \mid \mathcal D_n]$.

\item[Step 2:] Simulate $\widetilde \bb_j$ from the posterior of the random effects $[\bb_j \mid T_j^* > t, \mathcal Y_j(t), \widetilde \bftheta]$.

\item[Step 3:] Simulate $\tilde y_j(u)$ from $[y_j(u) \mid \widetilde \bb_j, \widetilde \bftheta]$.

\item[Step 4:] Simulate $\ddot \bb_j$ and $\{\breve \bb_j^{(\ell)}, \ell = 1, \ldots, L\}$ independently from the posterior of the random effects $[\bb_j \mid T_j^* > t, \{\mathcal Y_j(t), \tilde y_j(u)\}, \ddot \bftheta]$ and $[\bb_j \mid T_j^* > u, \{\mathcal Y_j(t), \tilde y_j(u)\}, \breve \bftheta^{(\ell)}]$, respectively.

\item[Step 5:] Simulate $\ddot T_j^*$ from $[T_j^* \mid T_j^* > u, \ddot \bb_j, \ddot \bftheta]$.

\item[Step 6:] If $\ddot T_j^* > u$, compute $\mbox{EKL}^{(q)}(u \mid t) = \log \Bigl \{ \displaystyle L^{-1} \sum \limits_{\ell = 1}^L \mathcal A_n^{(\ell)} \big / \mathcal A_d^{(\ell)} \Bigr \}$, where
\[
\mathcal A_n^{(\ell)} = h_j \bigl ( \ddot T_j^* \mid \mathcal H_j \bigl (\ddot T_j^*, \breve \bb_j^{(\ell)}, \breve \bftheta^{(\ell)}\bigr) \bigr ) \exp \biggl \{ - \displaystyle \int_0^{\ddot T_j^*} h_j \bigl ( s \mid \mathcal H_j \bigl (s, \breve \bb_j^{(\ell)}, \breve \bftheta^{(\ell)} \bigr) \bigr ) \; ds\biggr \},
\]
and
\[
\mathcal A_d^{(\ell)} = \exp \biggl \{ - \displaystyle \int_0^u h_j \bigl ( s \mid \mathcal H_j \bigl (s, \breve \bb_j^{(\ell)}, \breve \bftheta^{(\ell)} \bigr) \bigr ) \; ds \biggr \};
\]
otherwise set $\mbox{EKL}^{(q)}(u \mid t) = 0$.
\end{itemize}
The sample mean over the Monte Carlo samples provides the estimate of $\mbox{EKL}(u \mid t)$, i.e., $\mbox{E}\widehat{\mbox{K}}\mbox{L}(u \mid t) = Q^{-1} \sum_{q = 1}^Q \mbox{EKL}^{(q)}(u \mid t)$. The posterior distribution of the random effects in Steps~2 and 4 is not of a known form, and therefore the realizations $\widetilde \bb_j$, $\ddot \bb_j$ and $\{\breve \bb_j^{(\ell)}, \ell = 1, \ldots, L\}$ are obtained using a Metropolis-Hastings algorithm, with proposal distribution a multivariate Student's-$t$ distribution with mean $\widehat \bb = \argmax_{\bb} \{ \log p(\bb \mid T_j^* > t, \mathcal Y_j(t), \widehat \bftheta) \}$, variance-covariance matrix $V = \bigl \{ - \partial^2 \log p(\bb \mid T_j^* > t, \mathcal Y_j(t), \widehat \bftheta) \big / \partial \bb^\top \partial \bb \big |_{b = \hat b} \bigr \}^{-1}$, and four degrees of freedom, where $\widehat \bftheta$ denotes the posterior means based on $[\bftheta \mid \mathcal D_n]$. Step~3 just entails simulating a realization from the mixed effects model. In Step~5 again the distribution $[T_j^* \mid T_j^* > t, \ddot \bb_j, \ddot \bftheta]$ is not of standard form, and hence we simulate the realization $\ddot T_j^*$ using the inversion method, i.e., first we simulate $v \sim U(0, 1)$, and then using a line-search method we find the $\ddot T_j^*$ for which $S_j(\ddot T_j^*, \ddot \bb_j, \ddot \bftheta) \big / S_j(t, \ddot \bb_j, \ddot \bftheta) = v$, with $S_j(\cdot)$ given by \eqref{Eq:SurvivalFun}. Finally, in Step~6 it is sufficient to compute the numerator of \eqref{Eq:KL} to obtain the optimal $u$. An estimate of $\pi_j(u \mid t)$ can be obtained from a separate Monte Carlo scheme \citep[see e.g.,][]{rizopoulos:11}
\[
\widehat \pi_j(u \mid t) = \frac{1}{G} \sum \limits_{g = 1}^G \frac{\exp \biggl \{ - \displaystyle \int_0^u h_j \bigl ( s \mid \mathcal H_j \bigl (s, \mathring \bb_j^{(g)}, \mathring \bftheta^{(g)} \bigr) \bigr ) \; ds \biggr \}}{\exp \biggl \{ - \displaystyle \int_0^t h_j \bigl ( s \mid \mathcal H_j \bigl (s, \mathring \bb_j^{(g)}, \mathring \bftheta^{(g)} \bigr) \bigr ) \; ds \biggr \}},
\]
with $\{\mathring \bftheta^{(g)}, g = 1, \ldots, G\}$ a sample from $[\bftheta \mid \mathcal D_n]$, and $\mathring \bb_j^{(g)}$ a sample from $[\bb_j \mid T_j^* > t, \mathcal Y_j(t), \mathring \bftheta^{(g)}]$. Based on the Monte Carlo sample of $\mbox{EKL}(u \mid t)$ a 95\% credible interval van be constructed using the corresponding Monte Carlo sample percentiles. Using this simulation scheme we take as optimal $u$ the value $\hat u = \argmax_{u \in \{u_1, \ldots, u_{\max}\}} \{\mbox{EKL}(u \mid t)\}$, subject to the constraint that $\widehat \pi_j(u \mid t) \geq \kappa$, with $u_1, \ldots, u_{\max}$ denoting a grid of values in $(t, t^{up}]$.


\section{Analysis of the Aortic Valve Dataset} \label{Sec:AoValv}
We return to the Aortic Valve dataset introduced in Section~\ref{Sec:Intro}. Our aim is to use the existing data to build joint models that can be utilized for planning the visiting patterns of future patients from the same population. In our study, a total of 77 (27\%) patients received a sub-coronary implantation (SI) and the remaining 208 patients a root replacement (RR). These patients were followed prospectively over time with annual telephone interviews and biennial standardized echocardiographic assessment of valve function until July 8, 2010. Echo examinations were scheduled at 6 months and 1 year postoperatively, and biennially thereafter, and at each examination, echocardiographic measurements were taken. By the end of follow-up, 1262 assessments have been recorded, with an average of 4.3 measurements per patient (s.d. 2.4 measurements), 59 (20.7\%) patients had died, and 73 (25.6\%) patients required a re-operation on the allograft. Here we are interested in the composite event re-operation or death, which was observed for 125 (43.9\%) patients.

We start by defining two sets of joint models that we will use for planning measurements of future patients. Each set considers a separate longitudinal outcome that measures the functioning of the aortic valve, namely, the square root transform of aortic gradient (continuous outcome) and the aortic regurgitation (binary outcome). Preliminary analysis using plots of the subject-specific trajectories of the two outcomes, and fitting linear mixed effects and mixed effects logistic regression models suggested nonlinear profiles for aortic gradient and linear evolutions for the logit ($\mbox{logit}(x) = \log \{x / (1 - x)\}$) of the probabilities of aortic regurgitation. Hence, we let $\by_{i,1}$ denote the continuous aortic gradient longitudinal outcome, and $\by_{i,2}$ the binary aortic regurgitation outcome and postulate the models
\begin{eqnarray*}
y_{i,1}(t) & = & (\beta_{0,1} + b_{i0,1}) + (\beta_{1,1} + b_{i1,1}) B_n(t, 1) + (\beta_{2,1} + b_{i2,1}) B_n(t, 2)\\
& & + \, \beta_{3,1} \mbox{\tt Age}_i + \beta_{4,1} \mbox{\tt Female}_i + \eps_i(t),
\end{eqnarray*}
where $\bB_n(t, \{1,2\})$ denotes the B-spline basis for a natural cubic spline with boundary knots at baseline and 19 years and one internal knot placed at 3.7 years (i.e., the median of the observed follow-up times), {\tt Female} denotes the dummy variable for females, and
\begin{eqnarray*}
\mbox{logit} \bigl [ \Pr \{ y_{i,2}(t) = 1 \} \bigr ] & = & (\beta_{0,2} + b_{i0,2}) + (\beta_{1,2} + b_{i1,2}) t + \beta_{2,2} \mbox{\tt Age}_i + \beta_{3,2} \mbox{\tt Female}_i.
\end{eqnarray*}
Both $\bb_{i,1}$ and $\bb_{i,2}$ are assumed to follow multivariate normal distributions with mean zero and covariance matrices $\bD_1$ and $\bD_2$, respectively. For the survival process we consider four relative risk models, each positing a different association structure between the two processes, namely:
\begin{eqnarray*}
M_{1,k}: && h_{i,k}(t) = h_{0,k}(t) \exp \bigl \{\gamma_{1,k} \mbox{\tt Age}_i + \gamma_{2,k} \mbox{\tt Female}_i + \alpha_{1,k} \eta_{i,k}(t) \bigl \},\\
M_{2,k}: && h_{i,k}(t) = h_{0,k}(t) \exp \bigl \{\gamma_{1,k} \mbox{\tt Age}_i + \gamma_{2,k} \mbox{\tt Female}_i + \alpha_{2,k} \eta_{i,k}'(t)  \bigl \},\\
M_{3,k}: && h_{i,k}(t) = h_{0,k}(t) \exp \bigl \{\gamma_{1,k} \mbox{\tt Age}_i + \gamma_{2,k} \mbox{\tt Female}_i + \alpha_{1,k} \eta_{i,k}(t) + \alpha_{2,k} \eta_{i,k}'(t)  \bigl \},\\
M_{4,k}: && h_{i,k}(t) = h_{0,k}(t) \exp \Bigl \{\gamma_{1,k} \mbox{\tt Age}_i + \gamma_{2,k} \mbox{\tt Female}_i + \alpha_{1,k} \int_0^t \eta_{i,k}(s) ds  \Bigl \},\\
M_{5,k}: && h_{i,k}(t) = h_{0,k}(t) \exp \bigl (\gamma_{1,k} \mbox{\tt Age}_i + \gamma_{2,k} \mbox{\tt Female}_i + \bfalpha_k^\top \bb_{i,k} \bigl ),
\end{eqnarray*}
where $k = 1, 2$ denotes the relative risk models corresponding to the two sets of joint models for aortic gradient and aortic regurgitation, respectively. The baseline hazard was approximated using penalized B-splines approach described in Section~\ref{Sec:JM}. For all parameters we took standard prior distributions \citep{ibrahim.et.al:01, brown.ibrahim:03, brown:09}. In particular, for the vector of fixed effects of the longitudinal submodel $\bfbeta$, the regression parameters of the survival model $\bfgamma$, the vector of spline coefficients for the baseline hazard $\bfgamma_{h_0}$, and for the association parameter $\alpha$ we used independent univariate diffuse normal priors. For the variance of the error terms $\sigma^2$ we take an inverse-Gamma prior, while for covariance matrices we assumed an inverse Wishart prior.

Tables~\ref{Tab:Res-Long} and \ref{Tab:Res-Surv} present the posterior means and 95\% credible intervals of the parameters of the joint models for aortic gradient and aortic regurgitation fitted to the Aortic Valve data.
\begin{table}[ht]%
\centering
\scriptsize
\begin{tabular}{lrrrrrrrr}
 \hline
 & ~\hfill$M_{1,1}$\hfill~ & ~\hfill$M_{2,1}$\hfill~ & ~\hfill$M_{3,1}$\hfill~ & ~\hfill$M_{4,1}$\hfill~ & ~\hfill$M_{5,1}$\hfill~\\
 & Value (95\% CI) & Value (95\% CI) & Value (95\% CI) & Value (95\% CI) & Value (95\% CI) \\
   \hline
Intercept & 3.67 (3.33; 4.00) & 3.66 (3.32; 4.01) & 3.68 (3.34; 4.03) & 3.67 (3.34; 4.02) & 3.67 (3.32; 4.02) \\
  B-spln1 & 3.28 (2.79; 3.83) & 3.37 (2.84; 3.89) & 3.31 (2.81; 3.84) & 3.26 (2.76; 3.75) & 3.34 (2.85; 3.83) \\
  B-spln2 & 2.70 (2.33; 3.13) & 2.77 (2.34; 3.21) & 2.72 (2.35; 3.13) & 2.68 (2.28; 3.08) & 2.72 (2.31; 3.15) \\
  {\tt Age} & -0.02 (-0.02; -0.01) & -0.02 (-0.02; -0.01) & -0.02 (-0.02; -0.01) & -0.02 (-0.02; -0.01) & -0.02 (-0.02; -0.01) \\
  {\tt Female} & 0.17 (-0.06; 0.40) & 0.17 (-0.06; 0.39) & 0.17 (-0.04; 0.41) & 0.18 (-0.05; 0.41) & 0.18 (-0.04; 0.41) \\
  $\sigma$ & 0.61 (0.58; 0.65) & 0.62 (0.58; 0.65) & 0.62 (0.58; 0.65) & 0.61 (0.58; 0.65) & 0.62 (0.58; 0.66) \\
  $d_{11}$ & 0.64 (0.48; 0.83) & 0.62 (0.47; 0.80) & 0.63 (0.48; 0.84) & 0.64 (0.47; 0.82) & 0.61 (0.46; 0.81) \\
  $d_{21}$ & -0.69 (-1.44; -0.06) & -0.51 (-1.19; 0.10) & -0.64 (-1.38; -0.04) & -0.69 (-1.42; -0.05) & -0.53 (-1.23; 0.12) \\
  $d_{31}$ & -0.43 (-1.05; 0.09) & -0.40 (-0.97; 0.15) & -0.43 (-1.00; 0.13) & -0.46 (-1.10; 0.11) & -0.34 (-0.96; 0.28) \\
  $d_{22}$ & 13.61 (9.66; 18.74) & 13.87 (9.79; 19.14) & 13.50 (9.76; 18.14) & 13.91 (9.85; 19.09) & 13.06 (9.26; 17.83) \\
  $d_{32}$ & 8.14 (4.40; 13.62) & 9.46 (5.84; 14.20) & 8.05 (4.55; 12.82) & 8.70 (4.61; 13.66) & 8.01 (4.09; 13.47) \\
  $d_{33}$ & 7.04 (3.11; 12.99) & 8.08 (4.62; 13.25) & 6.63 (3.32; 11.71) & 7.80 (3.55; 14.06) & 7.31 (3.06; 15.57) \\
   \hline
  & ~\hfill$M_{1,2}$\hfill~ & ~\hfill$M_{2,2}$\hfill~ & ~\hfill$M_{3,2}$\hfill~ & ~\hfill$M_{4,2}$\hfill~ & ~\hfill$M_{5,2}$\hfill~\\
 & Value (95\% CI) & Value (95\% CI) & Value (95\% CI) & Value (95\% CI) & Value (95\% CI) \\
  \hline
Intercept & -2.01 (-3.74; -0.34) & -2.03 (-3.80; -0.41) & -1.89 (-3.43; -0.32) & -2.03 (-3.64; -0.43) & -1.98 (-3.57; -0.32) \\
  time & 0.38 (0.28; 0.48) & 0.39 (0.29; 0.49) & 0.39 (0.29; 0.50) & 0.37 (0.27; 0.47) & 0.40 (0.29; 0.49) \\
  {\tt Age} & -0.03 (-0.07; -0.00) & -0.03 (-0.07; 0.00) & -0.03 (-0.07; -0.00) & -0.03 (-0.07; -0.00) & -0.03 (-0.07; -0.00) \\
  {\tt Female} & 1.40 (0.29; 2.55) & 1.40 (0.28; 2.56) & 1.26 (0.23; 2.24) & 1.43 (0.31; 2.51) & 1.35 (0.25; 2.44) \\
  $d_{11}$ & 14.29 (8.77; 22.97) & 14.04 (9.16; 21.57) & 11.01 (7.19; 16.10) & 13.39 (8.18; 20.15) & 12.64 (8.08; 18.83) \\
  $d_{21}$ & -0.56 (-1.47; 0.17) & -0.38 (-1.11; 0.24) & 0.06 (-0.58; 0.63) & -0.48 (-1.31; 0.22) & -0.18 (-0.92; 0.46) \\
  $d_{22}$ & 0.47 (0.27; 0.80) & 0.44 (0.24; 0.73) & 0.48 (0.27; 0.84) & 0.52 (0.30; 0.90) & 0.43 (0.25; 0.74) \\
  \hline
\end{tabular}
\caption{Estimated coefficients and 95\% credibility intervals for the parameters of the longitudinal submodels fitted to the Aortic Valve dataset. The top part of the table refers to the results for aortic gradient and the bottom part for aortic regurgitation. $d_{ij}$ denotes the $ij$-th element of the corresponding covariance matrix of the random effects.}
\label{Tab:Res-Long}
\end{table}
\begin{table}[ht]%
\centering
\scriptsize
\begin{tabular}{lrrrrr}
  \hline
 & ~\hfill$M_{1,1}$\hfill~ & ~\hfill$M_{2,1}$\hfill~ & ~\hfill$M_{3,1}$\hfill~ & ~\hfill$M_{4,1}$\hfill~ & ~\hfill$M_{5,1}$\hfill~\\
 & Value (95\% CI) & Value (95\% CI) & Value (95\% CI) & Value (95\% CI) & Value (95\% CI) \\
  \hline
{\tt Age} & 0.02 (0.01; 0.04) & 0.02 (0.01; 0.04) & 0.02 (0.01; 0.04) & 0.02 (0.00; 0.03) & 0.02 (0.01; 0.04) \\
  {\tt Female} & -0.09 (-0.48; 0.29) & -0.02 (-0.43; 0.36) & -0.09 (-0.49; 0.30) & -0.10 (-0.51; 0.29) & 0.01 (-0.43; 0.45) \\
  $\alpha_1$ & 0.19 (0.09; 0.30) &  & 0.16 (0.02; 0.30) & 0.02 (0.01; 0.03) & -0.03 (-0.59; 0.51) \\
  $\alpha_2$ &  & 2.40 (1.07; 3.82) & 0.78 (-1.76; 2.99) &  & 0.43 (0.26; 0.65) \\
  $\alpha_3$ &  &  &  &  & -0.02 (-0.75; 0.81) \\
   \hline
& ~\hfill$M_{1,2}$\hfill~ & ~\hfill$M_{2,2}$\hfill~ & ~\hfill$M_{3,2}$\hfill~ & ~\hfill$M_{4,2}$\hfill~ & ~\hfill$M_{5,2}$\hfill~\\
& Value (95\% CI) & Value (95\% CI) & Value (95\% CI) & Value (95\% CI) & Value (95\% CI) \\
  \hline
{\tt Age} & 0.01 (-0.00; 0.03) & 0.01 (-0.00; 0.03) & 0.01 (-0.00; 0.03) & 0.01 (-0.00; 0.02) & 0.02 (0.00; 0.03) \\
  {\tt Female} & -0.02 (-0.43; 0.36) & -0.01 (-0.44; 0.40) & 0.07 (-0.40; 0.52) & 0.00 (-0.40; 0.41) & -0.02 (-0.51; 0.43) \\
  $\alpha_1$ & 0.03 (-0.00; 0.07) &  & -0.09 (-0.18; -0.01) & 0.001 (-0.002; 0.004) & -0.06 (-0.22; 0.04) \\
  $\alpha_2$ &  & 1.13 (0.32; 2.32) & 2.31 (0.84; 4.17) &  & 1.46 (0.43; 3.63) \\
   \hline
\end{tabular}
\caption{Estimated coefficients and 95\% credibility intervals for the parameters of the survival submodels fitted to the Aortic Valve dataset. The top part of the table refers to the results for aortic gradient and the bottom part for aortic regurgitation.}
\label{Tab:Res-Surv}
\end{table}%
With respect to the parameters of the longitudinal component we do not observe great differences between the fitted joint models for both aortic gradient and aortic regurgitation. For the relative risk submodels and for aortic gradient we see that when the terms $\eta_{i,1}(t)$ and $\eta_{i,1}'(t)$ are independently included in the model both of them show to be related with the risk of re-operation/death, however, when both of them are included in the model, the slope term $\eta_{i,1}'(t)$ does not seem to be associated with the hazard of the event anymore. Moreover, the area under the subject-specific longitudinal profile up to time $t$ is also strongly related to the risk for event at same time point, indicating that not only features of the trajectory at time $t$ are important but also at previous time points. For aortic regurgitation we observe that when combined, both the current value and the current slope term are associated with the risk of the composite event, whereas past values (i.e., the cumulative effect) do not seem to contribute in predicting future events.

We continue by investigating the performance of the fitted joint models with respect to predicting future events after different follow-up times. Table~\ref{Tab:Res-DICs} shows the DIC and the value of $\dclHat(t) \times n_t$ for $t = 5, 7, 9, 11$ and $13$.
\begin{table}[ht]
\centering
\scriptsize
\begin{tabular}{lrrrrr}
  \hline
 & ~\hfill$M_{1,1}$\hfill~ & ~\hfill$M_{2,1}$\hfill~ & ~\hfill$M_{3,1}$\hfill~ & ~\hfill$M_{4,1}$\hfill~ & ~\hfill$M_{5,1}$\hfill~\\
  \hline
DIC & 7237.26 & 7186.18 & 7195.57 & 7268.45 & 7186.34 \\
  $\dclHat(t = 5) \times n_t$ & -387.67 & -348.96 & -380.82 & -441.69 & -356.31 \\
  $\dclHat(t = 7) \times n_t$ & -342.75 & -309.69 & -336.95 & -391.78 & -315.26 \\
  $\dclHat(t = 9) \times n_t$ & -289.11 & -260.44 & -284.13 & -332.06 & -264.95 \\
  $\dclHat(t = 11) \times n_t$ & -233.56 & -208.11 & -229.27 & -270.28 & -212.06 \\
  $\dclHat(t = 13) \times n_t$ & -177.10 & -156.88 & -173.58 & -206.40 & -160.03 \\
   \hline
& ~\hfill$M_{1,2}$\hfill~ & ~\hfill$M_{2,2}$\hfill~ & ~\hfill$M_{3,2}$\hfill~ & ~\hfill$M_{4,2}$\hfill~ & ~\hfill$M_{5,2}$\hfill~\\
     \hline
DIC & 4166.76 & 4114.27 & 4049.99 & 4186.72 & 4061.53 \\
  $\dclHat(t = 5) \times n_t$ & -370.46 & -362.97 & -328.10 & -358.95 & -365.31 \\
  $\dclHat(t = 7) \times n_t$ & -328.84 & -321.25 & -288.47 & -318.74 & -322.70 \\
  $\dclHat(t = 9) \times n_t$ & -278.23 & -272.55 & -244.73 & -269.81 & -273.23 \\
  $\dclHat(t = 11) \times n_t$ & -223.45 & -218.67 & -195.18 & -216.21 & -219.19 \\
  $\dclHat(t = 13) \times n_t$ & -170.24 & -167.14 & -147.95 & -164.21 & -167.95 \\
   \hline
\end{tabular}
\caption{DIC and $\dclHat(t)$ evaluated at five time points for the fitted joint models.}
\label{Tab:Res-DICs}
\end{table}
From the set of joint models fitted with aortic gradient as the biomarker we observe that models $M_{2,1}$ and $M_{5,1}$ have almost identical DIC values but according to $\dclHat(t)$ model $M_{2,1}$ provides the best predictions for future events. For aortic regurgitation model $M_{3,2}$ is selected as the best model according to both DIC and $\dclHat(t)$. However, as a side note we observe that model $M_{5,2}$ has a lower DIC value than $M_{4,2}$ but the $\dclHat(t)$ was higher for model $M_{4,2}$ for all $t$, which again suggests that the DIC, as a global measure, may not always select the model that more accurately predicts future events after a particular time point $t$.

\begin{figure}[!h]%
\includegraphics[width = \textwidth]{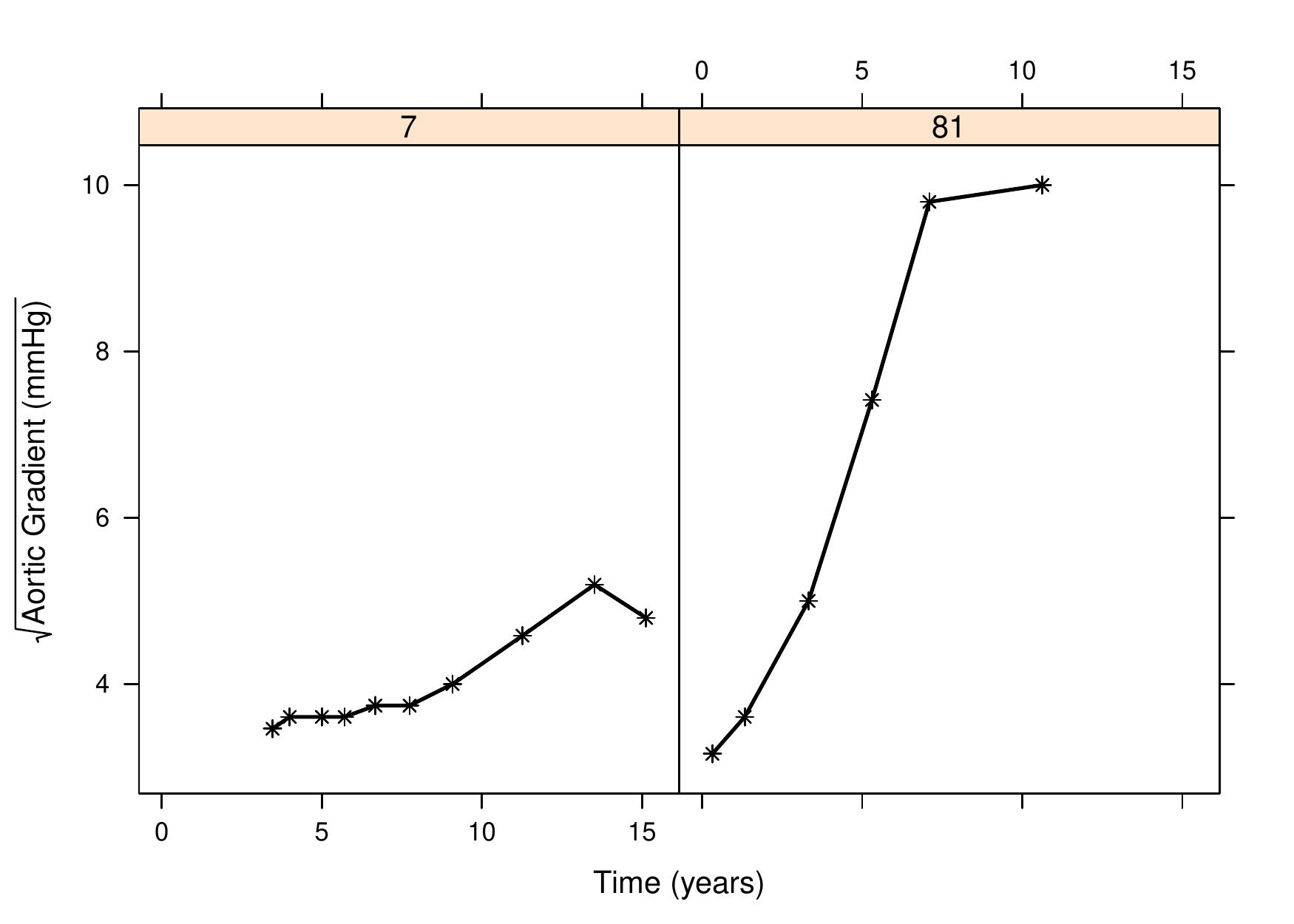}
\caption{Longitudinal trajectories of the square root aortic gradient for Patients~7 and 81.} \label{Fig:PatProfiles}
\end{figure}

To illustrate the use of the methodology developed in Section~\ref{Sec:Info} for selecting the next screening time point, we focus on the continuous marker and two patients from the Aortic Valve dataset, namely Patient~81 who is a male of 39.8 years old, and Patient~7 who is a male of 46.8 years old and neither had an event. These patients have been explicitly selected because of the shape of their longitudinal trajectories of the square root aortic gradient, shown in Figure~\ref{Fig:PatProfiles}.
More specifically, Patient~7 exhibits a slightly increasing trajectory of aortic gradient indicating that he remained in relatively good condition during the follow-up, whereas Patient~81 showed an increasing profile of aortic gradient indicative of a deterioration of his condition. We should note that in accordance with a realistic use of \eqref{Eq:Util} for planning the next measurement, the joint models presented above have been fitted in the version of the Aortic Valve dataset that excluded those two patients. Tables~\ref{Tab:Res-Info81} and \ref{Tab:Res-Info7} show the values of $\mbox{EKL}(u \mid t)$ (i.e., the first term of \eqref{Eq:Util}), and of $\pi_j(u \mid t)$ (i.e., the second term of \eqref{Eq:Util}) for these two patients computed in a dynamic manner, namely, after each of their longitudinal measurements (i.e., time point $t$). The upper limit $t^{up}$ of the interval $(t, t^{up}]$ was set as the minimum of $t^{max} = 5$ years and the time point $u$ for which $\pi_j(u \mid t)$ was equal to $\kappa = 0.8$. Then the optimal $u$ was taken as the point that maximizes $\mbox{EKL}(u \mid t)$ in a grid of five equidistant values in the interval $(t, t^{up}]$. All calculations have been based on joint model $M_{2,1}$.
\begin{table}[ht]%
\scriptsize
\centering
\begin{tabular}{cccccc}
  \hline
$t$ & $t^{up} - t$ & $u$ & $\mbox{EKL}(u \mid t)$ & $\pi_j(u \mid t)$ &   \\
  \hline
0.3 & 5 & 1.3 & -2.488 & 0.989 &   \\
   &  & 2.3 & -2.205 & 0.977 &   \\
   &  & 3.3 & -2.227 & 0.963 &   \\
   &  & 4.3 & -2.197 & 0.948 &   \\
   &  & 5.3 & -2.153 & 0.933 & * \\
   \hline
1.3 & 5 & 2.3 & -2.176 & 0.987 &   \\
   &  & 3.3 & -2.127 & 0.973 &   \\
   &  & 4.3 & -1.963 & 0.957 &   \\
   &  & 5.3 & -2.088 & 0.940 &   \\
   &  & 6.3 & -1.868 & 0.922 & * \\
   \hline
3.3 & 5 & 4.3 & -2.143 & 0.978 &   \\
   &  & 5.3 & -1.436 & 0.956 & * \\
   &  & 6.3 & -2.015 & 0.932 &   \\
   &  & 7.3 & -1.913 & 0.905 &   \\
   &  & 8.3 & -1.634 & 0.875 &   \\
   \hline
5.3 & 3.9 & 6.1 & -2.143 & 0.968 &   \\
   &  & 6.9 & -1.603 & 0.935 & * \\
   &  & 7.6 & -1.879 & 0.900 &   \\
   &  & 8.4 & -1.967 & 0.864 &  \\
   &  & 9.2 & -1.750 & 0.827 &   \\
   \hline
7.1 & 2.1 & 7.5 & -1.888 & 0.969 & * \\
   &  & 7.9 & -1.993 & 0.937 &   \\
   &  & 8.4 & -1.963 & 0.906 &   \\
   &  & 8.8 & -1.901 & 0.875 &   \\
   &  & 9.2 & -2.104 & 0.844 &  \\
   \hline
10.6 & 1.7 & 11.0 & -0.243 & 0.970 &   \\
   &  & 11.3 & -0.156 & 0.940 & * \\
   &  & 11.6 & -1.397 & 0.910 &   \\
   &  & 12.0 & -0.542 & 0.880 &   \\
   &  & 12.3 & -2.184 & 0.849 &   \\
   \hline
\end{tabular}
\caption{The values of $\mbox{EKL}(u \mid t)$ (i.e., the first term of \eqref{Eq:Util}), $\pi_j(u \mid t)$ (i.e., the second term of \eqref{Eq:Util}), for Patient~81 computed in a dynamic manner, after each of his longitudinal measurements (i.e., time point $t$), and for $u$ set as a regular sequence from $t$ to $t + 3$ of length five. The last column denotes which time point is selected to plan the next measurement.}
\label{Tab:Res-Info81}
\end{table}
\begin{table}[ht]%
\scriptsize
\centering
\begin{tabular}{cccccc}
  \hline
$t$ & $t^{up} - t$ & $u$ & $\mbox{EKL}(u \mid t)$ & $\pi_j(u \mid t)$ &   \\
  \hline
3.5 & 5 & 4.5 & -2.363 & 0.983 &   \\
   &  & 5.5 & -2.354 & 0.964 &   \\
   &  & 6.5 & -2.306 & 0.943 &   \\
   &  & 7.5 & -2.279 & 0.919 &   \\
   &  & 8.5 & -2.065 & 0.892 & * \\
   \hline
4 & 5 & 5.0 & -2.366 & 0.982 &   \\
   &  & 6.0 & -2.422 & 0.963 &   \\
   &  & 7.0 & -2.210 & 0.941 &   \\
   &  & 8.0 & -2.214 & 0.915 &   \\
   &  & 9.0 & -2.106 & 0.886 & * \\
   \hline
5 & 5 & 6.0 & -2.416 & 0.981 &   \\
   &  & 7.0 & -2.470 & 0.959 &   \\
   &  & 8.0 & -2.261 & 0.933 &   \\
   &  & 9.0 & -2.300 & 0.904 &   \\
   &  & 10.0 & -2.146 & 0.872 & * \\
   \hline
5.7 & 5 & 6.7 & -2.455 & 0.979 &   \\
   &  & 7.7 & -2.410 & 0.956 &   \\
   &  & 8.7 & -2.311 & 0.928 &   \\
   &  & 9.7 & -2.352 & 0.897 &   \\
   &  & 10.7 & -2.207 & 0.861 & * \\
   \hline
6.7 & 5 & 7.7 & -2.426 & 0.976 &   \\
   &  & 8.7 & -2.389 & 0.949 &   \\
   &  & 9.7 & -2.328 & 0.917 &   \\
   &  & 10.7 & -2.304 & 0.882 &   \\
   &  & 11.7 & -2.235 & 0.841 & * \\
   \hline
7.7 & 5 & 8.7 & -2.403 & 0.973 &   \\
   &  & 9.7 & -2.394 & 0.941 &   \\
   &  & 10.7 & -2.307 & 0.906 &   \\
   &  & 11.7 & -2.272 & 0.865 &   \\
   &  & 12.7 & -2.171 & 0.819 & * \\
   \hline
9.1 & 4.3 & 9.9 & -2.355 & 0.971 &   \\
   &  & 10.8 & -2.335 & 0.939 &   \\
   &  & 11.7 & -2.201 & 0.903 &   \\
   &  & 12.5 & -2.130 & 0.863 &   \\
   &  & 13.4 & -2.096 & 0.817 & * \\
   \hline
11.3 & 3.1 & 11.9 & -2.202 & 0.969 &   \\
   &  & 12.5 & -2.044 & 0.935 &   \\
   &  & 13.1 & -2.003 & 0.897 &   \\
   &  & 13.8 & -1.908 & 0.856 & * \\
   &  & 14.4 & -1.920 & 0.810 &   \\
   \hline
13.5 & 1.9 & 13.9 & -1.919 & 0.968 &   \\
   &  & 14.3 & -1.855 & 0.933 &   \\
   &  & 14.7 & -1.742 & 0.897 &   \\
   &  & 15.1 & -1.655 & 0.858 & * \\
   &  & 15.4 & -1.719 & 0.817 &   \\
   \hline
15.1 & 1.4 & 15.4 & -1.769 & 0.968 &   \\
   &  & 15.7 & -1.731 & 0.934 &   \\
   &  & 15.9 & -1.632 & 0.899 &   \\
   &  & 16.2 & -1.560 & 0.862 &   \\
   &  & 16.5 & -1.536 & 0.824 & * \\
   \hline
\end{tabular}
\caption{The values of $\mbox{EKL}(u \mid t)$ (i.e., the first term of \eqref{Eq:Util}), $\pi_j(u \mid t)$ (i.e., the second term of \eqref{Eq:Util}), for Patient~7 computed in a dynamic manner, after each of his longitudinal measurements (i.e., time point $t$), and for $u$ set as a regular sequence from $t$ to $t + 3$ of length five. The last column denotes which time point is selected to plan the next measurement.}
\label{Tab:Res-Info7}
\end{table}%
For Patient~7, who remained relatively stable during the first period of his follow-up, we observe that $\mbox{EKL}(u \mid t)$ always suggests waiting the full five years up to measurement seven; however, after the seventh measurement and because his profile started to slightly increase and his relative older age, the length of the interval $(t, t^{up}]$ for finding the optimal $u$ decreases, which in turn suggests to wait after each visit less time before requesting the next echocardiography. For Patient~81 on the other hand we observe that for the first two visits, where he exhibited relatively low levels of aortic gradient, $\mbox{EKL}(u \mid t)$ suggests to wait for a longer period years before requesting the next measurement. In accordance, $t^{up}$ is set to five years because the probability that he will survive more than five years was higher than $\kappa = 0.8$. After the third measurement however, when it starts to become clear that the condition of this patient deteriorates, $\mbox{EKL}(u \mid t)$ recommends to wait less time to request the next echocardiography. Accordingly from the fourth measurement onwards the length of the interval $(t, t^{up}]$ decreases and in the last measurement is set to 1.7 years that also indicates the deterioration of the patient.


\section{Discussion} \label{Sec:Disc}
In this paper we have presented two measures based on information theory that can be used to dynamically select models in time and optimally schedule longitudinal biomarker measurements. Contrary to standard screening procedures that require all patients to adhere to the same screening intervals, the combination of the framework of joint models for longitudinal and time-to-event data with these two measures allows one to tailor screening to the needs of individual patients and dynamically adapt during follow-up.

Motivated by the aim of the Aortic Valve dataset, in this work we have concentrated on a single endpoint. Nevertheless, using recent advances in the joint modeling literature \citep{andrinopoulou.et.al:14, dantan.et.al:11} the same ideas can be relatively easily extended to multiple endpoints (competing risks) or intermediate states combined with one or several endpoints (multistate settings). In such settings the first term of \eqref{Eq:Util} will remain the same, but the second term $\pi_j(u \mid t)$ will need to be replaced by the corresponding cumulative incidence function. Furthermore, our approach also could be combined with concepts, such as the value of information \citep{koerkamp.et.al:06} to incorporate the relative costs of measuring the biomarker.

Finally, to facilitate the computation of $\dclHat(t)$ and $\mbox{E}\widehat{\mbox{K}}\mbox{L}$ has been implemented in functions \texttt{cvDCL()} and \texttt{dynInfo()}, respectively, available in package the \textbf{JMbayes} (version 0.7-0; \citealp{JMbayes}) for the \R\, programming language (freely available from the Comprehensive \R\, Archive Network at \url{http://cran.r-project.org/package=JMbayes}). A working example of how these functions should be used can be found in the supplementary material.


\bibliographystyle{asa}
\bibliography{NextVisit}
\end{document}